\documentclass[a4paper,12pt]{article}
\usepackage{amssymb,amsmath}
\usepackage{fancyhdr,verbatim}

\usepackage{a4}
\usepackage{parskip}
\usepackage{appendix}
\usepackage{color,soul,rotating}
\usepackage{amsthm,amsxtra,amscd,relsize,multirow}
\usepackage[all,2cell]{xy}\UseAllTwocells

\newcommand{\sect}[1]{\setcounter{equation}{0}\section{#1}}

\def\cosh{\mathrm{cosh}}

\def\axs{AdS_5\times S^5}
\newcommand{\eq}[1]{\begin{equation} #1 \end{equation}}
\newcommand{\al}[1]{\begin{align} #1 \end{align}}

\begin{document}
\begin{titlepage}
\title{Three-point functions of semiclassical string states and conserved currents in $AdS_4\times CP^3$}
\author{D.~Arnaudov\thanks{E-mail address: dlarnaudov@phys.uni-sofia.bg}
\ \\ \ \\
Department of Physics, Sofia University,\\
5 J. Bourchier Blvd, 1164 Sofia, Bulgaria
}
\date{}
\end{titlepage}

\maketitle

\begin{abstract}
In this paper we study the three-point correlation functions of two scalar operators with large conformal dimensions and the $R$-current or stress-energy tensor at strong coupling with the help of the AdS$_4$/CFT$_3$ correspondence. The scalar operators are dual to semiclassical strings in $AdS_4\times CP^3$, which are point-like in AdS. We establish thorough concordance between string theory results at large coupling constant and general predictions coming from Ward identities in the dual three-dimensional superconformal gauge theory.
\end{abstract}

\sect{Introduction}

One of the active areas of research in theoretical high-energy physics in recent years has been the correspondence between gauge and string theories. Following the impressive conjecture made by Maldacena \cite{Maldacena} that type IIB string theory on $AdS_5\times S^5$ is dual to ${\cal N}=4$ super-Yang--Mills theory with a large number of colors, an explicit realization of the AdS/CFT correspondence was provided in \cite{GKP}. After that many convincing results have been achieved, paving the way for the subject to become an indispensable tool in probing such diverse areas as the dynamics of quark-gluon plasma and high-temperature superconductivity.

A key feature of the duality is the connection between planar correlation functions of conformal primary operators in the gauge theory and correlators of corresponding vertex operators of closed strings with $S^2$ worldsheet topology. Recently, some progress was accomplished in the study of three- and four-point correlation functions with two and three vertex operators with large quantum numbers at strong coupling. The remaining operators were chosen to be various supergravity states with quantum numbers and dimensions of order one. It was shown that the large $\sqrt{\lambda}$ behavior of such correlators is fixed by a semiclassical string trajectory governed by the heavy operator insertions, and with sources provided by the vertex operators of the light states.

Initially this approach was utilized in the computation of two-point functions of heavy operators in \cite{Polyakov:2002}--\cite{Buchbinder:2010vw}. More recently the above procedure was extended to certain three-point correlators in \cite{Janik:2010gc,Zarembo:2010,Costa:2010}. A method based on heavy vertex operators was proposed in \cite{Roiban:2010}. Further developments in the calculation of correlators with two string states are presented in~\cite{Hernandez:2010}. The main goal of these investigations is elucidation of the structure of three-point functions of three semiclassical operators \cite{Klose:2011}.

Inspired by these studies, in the present paper we consider correlation functions of two massive states with large charges and a conserved current ($R$-current or stress-energy tensor) in the bosonic sector of ABJM theory from the point of view of strings in $AdS_4\times CP^3$. Our approach is mostly based on the previous works of \cite{Zarembo:2010,Georgiou:2013}. We also check the validity of our results by comparison with related field theory Ward identities.

The paper is organized as follows. In the next section we give a detailed derivation of the three-point function of two scalar operators with large charges and the $R$-current. The relevant structure constant complies with a Ward identity calculation. In the third section we present the three-point correlator of two arbitrary string states and the stress-energy tensor. In the conclusion we discuss the results and make some general remarks.

\sect{Three-point correlator of two string states and $R$-current}

The correlation functions with $R$-symmetry current have attracted the interest of researchers of the AdS/CFT correspondence from the very inception of the duality. The correlator of three $R$-current states was calculated at strong coupling via type-IIB supergravity, and the result was in concordance with field theory results \cite{Freedman:1998tz,Chalmers:1998xr}. Correlation functions of $R$-symmetry current and two BPS operators were also studied \cite{Freedman:1998tz}, and once more there was a match between supergravity and field-theoretic findings with the help of particular Ward identities.

The cases considered before dealt only with BPS states, because one could work only in the supergravity approximation. In this section we calculate the three-point correlator of an $R$-current candidate along with two semiclassical operators corresponding to string solutions. Our results at large 't Hooft coupling are perfectly compatible with a Ward identity in the dual conformal theory, which is yet another successful confirmation of the AdS/CFT correspondence.

\subsection{Form of correlator and Ward identity}
We adopt the following conventions: capital letters like $X^M,\,M=0,\dots,9$ denote ten-dimensional coordinates, while the lowercase, $x$ for example, describe four-dimensional ones and points in the dual three-dimensional field theory. Unless stated otherwise lengths between points $|x-y|$ are assumed to be for the three-dimensional theory. Greek letters $\mu=0,\dots,3$ will serve as indices of AdS directions, while Latin ones $i=0,1,2$ will denote boundary ones. Moreover, we will work in the Euclidean continuation of $AdS_4$.

The general form of the three-point function of a scalar operator ${\cal O}_{\Delta}$ with conformal dimension $\Delta$, its conjugate $\bar{\cal O}_{\Delta}$ and a vector $V_i$ is completely fixed by conformal symmetry. It is given in three dimensions by \cite{Fradkin:1998}
\al{\label{vec}
&\langle V_i(x)\,{\cal O}_{\Delta}(x_1)\,\bar{\cal O}_{\Delta}(x_2)\rangle
=\frac{C_{123}(\lambda)}{x_{12}^{2\Delta-1}|x_1-x|\,|x_2-x|}E_i^x(x_1,x_2)\,,\\
&x_{12}\equiv|x_1-x_2|\,,\qquad E_i^x(x_1,x_2)=\frac{(x_1-x)_i}{(x_1-x)^2}-\frac{(x_2-x)_i}{(x_2-x)^2}\,.
\nonumber
}
We are interested in the case of $V_i$ being the $R$-symmetry current $j_i^R$. We also assume that ${\cal O}_{\Delta}$ has $R$-charge, which leads to the following Ward identity for \eqref{vec}
\al{\nonumber
\langle\partial^ij_i^R(x)\,{\cal O}_{\Delta}(x_1)\,\bar{\cal O}_{\Delta}(x_2)\rangle
&=\delta^3(x_1-x)\langle\delta{\cal O}_{\Delta}(x_1)\,\bar{\cal O}_{\Delta}(x_2)\rangle\,+\\
\delta^3(x_2-x)\langle{\cal O}_{\Delta}(x_1)\,\delta\bar{\cal O}_{\Delta}(x_2)\rangle
&=J[\delta^3(x_1-x)-\delta^3(x_2-x)]\langle{\cal O}_{\Delta}(x_1)\,\bar{\cal O}_{\Delta}(x_2)\rangle\,,
\label{W-R}
}
where $J$ is the $R$-charge of ${\cal O}_{\Delta}$ and we assume that two-point functions are unit normalized. We take the derivative of \eqref{vec} and obtain
\eq{
\partial^i\langle V_i(x)\,{\cal O}_{\Delta}(x_1)\,\bar{\cal O}_{\Delta}(x_2)\rangle
=-4\pi C_{123}(\lambda)[\delta^3(x_1-x)-\delta^3(x_2-x)]\frac{1}{x_{12}^{2\Delta}}\,.
\label{diff-vec}
}
Combining \eqref{W-R} and \eqref{diff-vec} one gets
\eq{
C_{123}(\lambda)=-\frac{J}{4\pi}\,.
\label{C-R}
}
This equation gives an all-loop expression for the fusion coefficient $C_{123}(\lambda)$. In the following subsection we will show that string theory provides correctly both the space-time behavior of the correlator \eqref{vec} and the expression in \eqref{C-R}.

\subsection{Holographic calculation of $\langle j^{\mu}_R(x)\,{\cal O}_{\Delta}(x_1)\,\bar{\cal O}_{\Delta}(x_2)\rangle$}
The present subsection is devoted to the computation of the three-point function of two heavy scalar operators and the $R$-current via the AdS/CFT correspondence. The scalars are dual to semiclassical strings, which are point-like in AdS and rotate in $CP^3$ in a general way. First, we need to find the bulk supergravity field that corresponds to the $R$-current. Gaining intuition from \cite{Kim:1985} one can deduce that the dual field includes the fluctuations of the bulk metric $h_{\mu a}$ with one $AdS_4$ index and a $CP^3$ index; together with various components of RR-potentials, which do not contribute to our results according to \cite{Zarembo:2010} because we are working in the leading semiclassical approximation, i.e. at strong coupling. The metric can be expanded in the components of $CP^3$ Killing vectors
\eq{
h_{\mu a}=\sum_IH_{\mu}^I(x)Y_a^I(\Omega)\,.
}
We choose the following representation of $CP^3$ in terms of angles $(\xi,\theta_1,\theta_2,\psi,\varphi_1,\varphi_2)$
\al{\nonumber
ds^2_{CP^3}&=d\xi^2+\cos^2\xi\sin^2\xi(d\psi+\frac12\cos\theta_1d\varphi_1-\frac12\cos\theta_2d\varphi_2)^2\\
&+\frac14\cos^2\xi(d\theta_1^2+\sin^2\theta_1d\varphi_1^2)+\frac14\sin^2\xi(d\theta_2^2+\sin^2\theta_2 d\varphi_2^2)\,.
}
Let us choose for concreteness dynamics only in the plane with $\psi$ as polar angle. Then the relevant Killing vector will be the corresponding $R$-current which acts as the generator of rotations in the plane. The Killing vector and its unique covariant component assume the form
\eq{
Y^{[\psi]}=\frac{\partial}{\partial\psi}\,,\qquad Y^{[\psi]}_{\psi}=\cos^2{\xi}\sin^2{\xi}\,.
}

Following \cite{Zarembo:2010} the three-point correlation function has the schematic evaluation
\eq{
\frac{\langle j^{[\psi]}_i(x)\,{\cal O}_{\Delta}(x_1)\,\bar{\cal O}_{\Delta}(x_2)\rangle}{\langle{\cal O}_{\Delta}(x_1)\,{\bar {\cal O}}_{\Delta}(x_2)\rangle}=\big\langle H^{[\psi]}_i(x,x_3=0)\,\frac{1}{{\cal Z}_{\rm str}}\int DX\,e^{-S_{\rm str}[X,\Phi]}\big\rangle_{\rm bulk}\,,
\label{corr-R}
}
where $\Phi$ labels all the relevant supergravity fields. The field $H^{[\psi]}(x,x_3=0)$ has much smaller conformal dimension than that of the string states. Therefore the string can be treated in semiclassical approximation via Polyakov action, while the supergravity fields should conform to the supergravity approximation, i.e. the string action can be expanded in powers of $h_{\mu a}$
\al{
S_{\rm str}&=\frac{\sqrt{\tilde\lambda}}{4\pi}\int d^2\sigma\,\sqrt{g}g^{\alpha\beta}\,\partial_{\alpha}X^M\partial_{\beta}X^NG_{MN}+{\rm fermions}\,,\\
\frac{\delta S_{\rm str}}{\delta h_{\mu a}(X)}&=\frac{\sqrt{\tilde\lambda}}{2\pi}\int d^2\sigma\,(\partial_{\tau}X^{\mu}\partial_{\tau}X^a+\partial_{\sigma}X^{\mu}\partial_{\sigma}X^a)\,.
\label{sstr}
}
The relation between $\tilde\lambda$ and the 't Hooft coupling $\lambda$ is $\tilde\lambda^2=32\pi^2\lambda^2$. After substituting \eqref{sstr} in \eqref{corr-R} and retaining only the linear in $h_{\mu a}$ term we obtain
\al{
&\frac{\langle j^{[\psi]}_i(x)\,{\cal O}_{\Delta}(x_1)\,\bar{\cal O}_{\Delta}(x_2)\rangle}{\langle{\cal O}_{\Delta}(x_1)\,{\bar {\cal O}}_{\Delta}(x_2)\rangle}=-\big\langle H_i(x,x_3=0)\frac{\delta S_{\rm str}[X,\Phi=0]}{\delta h_{\mu a}(Z)}h_{\mu a}(Z)\big\rangle_{\rm bulk}=\nonumber\\
&-\frac{\sqrt{\tilde\lambda}}{2\pi}\int d^2\sigma\,(\partial_{\tau}Z^{\mu}\partial_{\tau}Z^a+\partial_{\sigma}Z^{\mu}\partial_{\sigma}Z^a)\langle
H_i(x,x_3=0)H_{\mu}(z)\rangle_{\rm bulk}\,Y_a^{[\psi]}(\Omega)\,.
\label{corr-R'}
}
At this point a few comments are in order. The path integral in \eqref{corr-R} is evaluated by saddle-point approximation on the classical solution describing a string that goes from the boundary and back. The expression $\langle H_i(x,x_3=0)H_{\mu}(z)\rangle_{\rm bulk}$ in \eqref{corr-R'} is actually the bulk-to-boundary propagator of two vector supergravity fields \cite{Freedman:1998tz}
\eq{
G_{\mu i}(z;x,x_3=0)=\frac{2}{\pi^2}\frac{z_3}{[z_3^2+(z-x)^2]^2}\!\left(\delta_{\mu i}-\frac{2(z-x)_{\mu}\,(z-x)_i}{z_3^2+(z-x)^2}\right).
}

In order to calculate the three-point function we need only to choose an appropriate solution to the string equations of motion dual to the massive states in the correlators. We are interested solely in the dynamics in AdS, because only it enters \eqref{corr-R'}. Moreover, we have already specified that the string is rotating along the $\psi$ direction of $CP^3$. In AdS the solution is purely point-like. It has the following form in Euclidean Poincare coordinates~\cite{Janik:2010gc}
\eq{
z_0=\frac{x_{12}}{2}\tanh(\kappa\tau)\,,\qquad z_3=\frac{x_{12}}{2\cosh(\kappa\tau)}\,,
\label{sol}
}
where we have assumed that the insertion points of the heavy operators are symmetric with respect to the origin of the coordinate system. Also, without loss of generality we have constrained the boundary heavy operators to lie on the Euclidean time direction: $x_{12}\equiv|x^0_1-x^0_2|$. We would like to point out that our considerations are applicable to generic string solutions with arbitrary charges in $CP^3$.

Substituting all above in \eqref{corr-R'}, we get for the three-point correlator
\al{
&\frac{\langle j^{[\psi]}_i(x)\,{\cal O}_{\Delta}(x_1)\,\bar{\cal O}_{\Delta}(x_2)\rangle}{\langle{\cal O}_{\Delta}(x_1)\,{\bar {\cal O}}_{\Delta}(x_2)\rangle}=
-\frac{2}{\pi^2}\!\left[\frac{\sqrt{\tilde\lambda}}{2\pi}\int_0^{2\pi}d\sigma\,\cos^2\xi\sin^2\xi\,\partial_{\tau}\psi\right]\!{\cal I}_i
=-\frac{2J}{\pi^2}\,{\cal I}_i\,,\\
&{\cal I}_i=\int_{-\infty}^{\infty}d\tau\,\frac{z_3\,\partial_{\tau}z_0}{[z_3^2+(z-x)^2]^2}\!\left(\delta_{0i}+\frac{2x_0(z-x)_i}{z_3^2+(z-x)^2}\right)
=\frac{\pi}{8}\,\frac{x_{12}\,E_i^x(x_1,x_2)}{|x_1-x|\,|x_2-x|}\,,
}
where we have used in the last equation that $x_1+x_2=0$. Finally we obtain for the three-point function
\eq{
\langle j^{[\psi]}_i(x)\,{\cal O}_{\Delta}(x_1)\,\bar{\cal O}_{\Delta}(x_2)\rangle=
-\frac{J}{4\pi}\,\frac{E_i^x(x_1,x_2)}{x_{12}^{2\Delta-1}|x_1-x|\,|x_2-x|}\,,
}
which exactly coincides with \eqref{vec} provided that
\eq{
C_{123}(\lambda\gg1)=-\frac{J}{4\pi}\,,
}
i.e. we also have complete agreement with the all-loop prediction \eqref{C-R}.

\sect{Three-point correlator with stress-energy tensor}

The present section is devoted to evaluation of the three-point function of two semiclassical scalar operators dual to string states and the stress-energy tensor. Correlation functions with the stress-energy tensor have been studied before, especially considering the conformal anomaly \cite{Osborn:1993cr}--\cite{Arutyunov:1999nw}.

\subsection{Space-time dependence of correlator and Ward identity}
The correlation function of any tensor $V_{ij}$ and two scalars with definite conformal dimension is almost completely fixed by conformal symmetry, excluding the structure constant
\al{\label{tens}
&\langle V_{ij}(x)\,{\cal O}_{\Delta}(x_1)\,\bar{\cal O}_{\Delta}(x_2)\rangle=\frac{C_{123}(\lambda)}{x_{12}^{2\Delta-1}|x_1-x|\,|x_2-x|}F_{ij}(x,x_1,x_2)\,,\\
&F_{ij}(x,x_1,x_2)=E_i^x(x_1,x_2)E_j^x(x_1,x_2)-\frac{\delta_{ij}}{3}\frac{x_{12}^2}{(x_1-x)^2(x_2-x)^2}\,.
\nonumber
}
When $V_{ij}=T_{ij}$, the following Ward identity concerning the conservation of $T_{ij}$ is satisfied
\eq{
\partial^i\langle T_{ij}(x)\,{\cal O}(x_1)\,\bar{\cal O}(x_2)\rangle=
\langle{\cal O}(x)\,\bar{\cal O}(x_1)\rangle\,\partial_j\delta^3(x-x_2)+\langle{\cal O}(x)\,\bar{\cal O}(x_2)\rangle\,\partial_j\delta^3(x-x_1)\,.
}
This equation can be integrated by $x$, which gives via Gauss's theorem
\eq{
C_{123}(\lambda)=-\frac{3}{8\pi}\,\Delta(\lambda)\,.
\label{C-T}
}
This result is an all-loop prediction which we are striving to reproduce at large coupling constant below.

\subsection{Calculation of $\langle T_{ij}(x)\,{\cal O}_{\Delta}(x_1)\,\bar{\cal O}_{\Delta}(x_2)\rangle$}
The present problem is solved analogously to the case in the previous section. The stress-energy tensor $T_{ij}$ is dual to the fluctuations of the metric $h_{\mu\nu}=g_{\mu\nu}-g_{\mu\nu}^{\rm AdS}$ of $AdS_4$. Consequently
\eq{
\langle T_{ij}(x)\,{\cal O}_{\Delta}(x_1)\,\bar{\cal O}_{\Delta}(x_2)\rangle=
\langle{\hat h}_{ij}(x)\,{\cal O}_{\Delta}(x_1)\,\bar{\cal O}_{\Delta}(x_2)\rangle\,.
}
We will need below the bulk-to-boundary propagator for gravitons, which can be extracted from the solution to the linearised equations of motion in de Donder gauge \cite{Liu:1998bu,Arutyunov:1999nw}
\eq{
h_{\mu\nu}(x,x_3)=\frac{8}{\pi^2}\int d^3y\,K(x,x_3;y)\,j_{\mu}^i(x-y)\,j^j_{\nu}(x-y)\,{\cal E}_{ij,kl}\,{\hat h}^{kl}(y)\,,
}
where
\eq{
K(x,x_3;y)=\frac{x_3^3}{[x_3^2+(x-y)^2]^3}\,,\quad j_{\mu}^i(x)=\delta_{\mu}^i-\frac{2x_{\mu}x^i}{x_3^2+x^2}\,,\quad
{\cal E}_{ij,kl}=\frac{\delta_{ik}\delta_{jl}+\delta_{il}\delta_{jk}}{2}-\frac{\delta_{ij}\delta_{kl}}{3}\,.
\nonumber
}
In order to calculate the three-point function at strong coupling we will proceed in a similar way as before
\al{
&\frac{\langle{\hat h}_{ij}(x)\,{\cal O}_{\Delta}(x_1)\,\bar{\cal O}_{\Delta}(x_2)\rangle}{\langle{\cal O}_{\Delta}(x_1)\,\bar{\cal O}_{\Delta}(x_2)\rangle}=
-\big\langle{\hat h}_{ij}(x,x_3=0)\,\frac{\delta S_{\rm str}[X,\Phi=0]}{\delta h^{\nu}_{\mu}(Z)}h^{\nu}_{\mu}(Z)\big\rangle_{\rm bulk}=\nonumber\\
&-\frac{\sqrt{\tilde\lambda}}{2\pi}\int d^2\sigma\,(\partial_{\tau}Z^{\mu}\partial_{\tau}Z_{\nu}+\partial_{\sigma}Z^{\mu}\partial_{\sigma}Z_{\nu })\,
\langle{\hat h}_{ij}(x,x_3=0)h^{\nu}_{\mu}(z)\rangle_{\rm bulk}\,.
}
The relevant string solution is again \eqref{sol} with arbitrary dynamics in $CP^3$. We obtain
\al{
&\frac{\langle{\hat h}_{ij}(x)\,{\cal O}_{\Delta}(x_1)\,\bar{\cal O}_{\Delta}(x_2)\rangle}{\langle{\cal O}_{\Delta}(x_1)\,\bar{\cal O}_{\Delta}(x_2)\rangle}=
-\frac{8\sqrt{\tilde\lambda}}{\pi^2}\int_{-\infty}^{\infty}d\tau\,\frac{z_3^3\,\partial_{\tau}Z^{\mu}\partial_{\tau}Z_{\nu}\,j_{\mu}^k(z-x)\,j_l^{\nu}(z-x)\,{\cal E}_{kl,ij}}{[z_3^2+(z-x)^2]^3}=\nonumber\\
&-\frac{8\sqrt{\tilde\lambda}}{\pi^2}\int_{-\infty}^{\infty}d\tau\,\frac{z_3\big(\partial_{\tau}z_0\big)^2}{[z_3^2+(z-x)^2]^3}\!
\left(\delta_{0i}\delta_{0j}-\frac{\delta_{ij}}{3}+\frac{2x_0\big[\delta_{0i}(z-x)_j+\delta_{0j}(z-x)_i-\frac{2\delta_{ij}(z-x)_0}{3}\big]}{z_3^2+(z-x)^2}\right.\nonumber\\
&\left.+\,\frac{4x_0^2[(z-x)_i(z-x)_j-\frac{\delta_{ij}}{3}(z-x)^2]}{[z_3^2+(z-x)^2]^2}\right),
}
where in the second line we have done all the necessary contractions using the explicit form of the solution \eqref{sol}. After tedious but straightforward calculations we get
\eq{
\frac{\langle{\hat h}_{ij}(x)\,{\cal O}_{\Delta}(x_1)\,\bar{\cal O}_{\Delta}(x_2)\rangle}{\langle{\cal O}_{\Delta}(x_1)\,\bar{\cal O}_{\Delta}(x_2)\rangle}=
-\frac{3\kappa\sqrt{\tilde\lambda}}{8\pi}\frac{x_{12}F_{ij}(x,x_1,x_2)}{|x_1-x|\,|x_2-x|}\,,
}
which looks exactly like \eqref{tens} if
\eq{
C_{123}(\lambda\gg1)=-\frac{3\kappa\sqrt{\tilde\lambda}}{8\pi}=-\frac{3E}{8\pi}\,,
}
where $E$ is the string energy. The last equation obviously conforms to the result \eqref{C-T} following from a Ward identity, because according to the AdS/CFT dictionary $E=\Delta(\lambda)$.

\sect{Conclusion}

The AdS/CFT correspondence was subject to many significant developments in recent years. One of the active areas of research has been the holographic calculation of three-point functions at strong coupling. The correlation function of three massive string states escapes full comprehension so far~\cite{Klose:2011}, but we have uncovered almost all features of correlators containing two heavy and one light states in the semiclassical approximation~\cite{Zarembo:2010,Costa:2010,Roiban:2010}.

The present paper continues this line of investigations by considering string theory on $AdS_4\times CP^3$ dual to the three-dimensional ABJM theory and computing leading three-point functions at large coupling constant, applying the ideas of \cite{Zarembo:2010}. We examine the method in the case of two scalar operators with large charges and a conserved current (either an $R$-symmetry current or the stress-energy tensor). We reproduce the correct space-time behavior of correlators and verify that the structure constants we have obtained at strong coupling are in perfect agreement with corresponding field theory derivations based on Ward identities. Our study extends the results presented in \cite{Georgiou:2013} to the AdS$_4$/CFT$_3$ case.

One of the future directions for exploration may be the connection of our work to recent developments in the calculation of correlators with heavy states based on integrability methods in ${\cal N}=4$ SYM \cite{Escobedo:2010} and, which is more relevant, ABJM theory \cite{Hirano:2012}.



\begin{thebibliography}{99}

\bibitem{Maldacena}
  J.~M.~Maldacena,
  ``The large $N$ limit of superconformal field theories and supergravity,''
  Adv. Theor. Math. Phys. {\bf 2}, 231 (1998)
  [hep-th/9711200].

\bibitem{GKP}
  S.~S.~Gubser, I.~R.~Klebanov and A.~M.~Polyakov,
  ``Gauge theory correlators from non-critical string theory,''
  Phys. Lett. B {\bf 428}, 105 (1998)
  [hep-th/9802109].
$\bullet$
  E.~Witten,
  ``Anti-de Sitter space and holography,''
  Adv. Theor. Math. Phys. {\bf 2}, 253 (1998)
  [hep-th/9802150].


\bibitem{Polyakov:2002}
  A.~M.~Polyakov,
  ``Gauge fields and space-time,''
  Int. J. Mod. Phys. A {\bf 17S1}, 119 (2002)
  [hep-th/0110196].

\bibitem{GKP2}
  S.~S.~Gubser, I.~R.~Klebanov and A.~M.~Polyakov,
  ``A semi-classical limit of the gauge/string correspondence,''
  Nucl. Phys. B {\bf 636}, 99 (2002)
  [hep-th/0204051].

\bibitem{Tseytlin:2003}
  A.~A.~Tseytlin,
  ``On semiclassical approximation and spinning string vertex operators in $\axs$,''
  Nucl. Phys. B {\bf 664}, 247 (2003)
  [hep-th/0304139].

\bibitem{Buchbinder:2010}
  E.~I.~Buchbinder,
  ``Energy-Spin Trajectories in $\axs$ from Semiclassical Vertex Operators,''
  JHEP {\bf 1004}, 107 (2010)
  [arXiv:1002.1716].

\bibitem{Buchbinder:2010vw}
  E.~I.~Buchbinder and A.~A.~Tseytlin,
  ``On semiclassical approximation for correlators of closed string vertex operators in AdS/CFT,''
  JHEP {\bf 1008}, 057 (2010)
  [arXiv:1005.4516].

\bibitem{Janik:2010gc}
  R.~A.~Janik, P.~Surowka and A.~Wereszczynski,
  ``On correlation functions of operators dual to classical spinning string states,''
  JHEP {\bf 1005}, 030 (2010)
  [arXiv:1002.4613].


\bibitem{Zarembo:2010}
  K.~Zarembo,
  ``Holographic three-point functions of semiclassical states,''
  JHEP {\bf 1009}, 030 (2010)
  [arXiv:1008.1059].

\bibitem{Costa:2010}
  M.~S.~Costa, R.~Monteiro, J.~E.~Santos and D.~Zoakos,
  ``On three-point correlation functions in the gauge/gravity duality,''
  JHEP {\bf 1011}, 141 (2010)
  [arXiv:1008.1070].

\bibitem{Roiban:2010}
  R.~Roiban and A.~A.~Tseytlin,
  ``On semiclassical computation of 3-point functions of closed string vertex operators in $\axs$,''
  Phys. Rev. D {\bf 82}, 106011 (2010)
  [arXiv:1008.4921].

\bibitem{Hernandez:2010}
  R.~Hern\'andez,
  ``Three-point correlation functions from semiclassical circular strings,''
  J. Phys. A {\bf 44}, 085403 (2011)
  [arXiv:1011.0408].
$\bullet$
  S.~Ryang,
  ``Correlators of Vertex Operators for Circular Strings with Winding Numbers in $\axs$,''
  JHEP {\bf 1101}, 092 (2011)
  [arXiv:1011.3573].
$\bullet$
  D.~Arnaudov and R.~C.~Rashkov,
  ``Semiclassical calculation of three-point functions in $AdS_4\times CP^3$,''
  Phys. Rev. D {\bf 83}, 066011 (2011)
  [arXiv:1011.4669].
$\bullet$
  G.~Georgiou,
  ``Two and three-point correlators of operators dual to folded string solutions at strong coupling,''
  JHEP {\bf 1102}, 046 (2011)
  [arXiv:1011.5181].
$\bullet$
  J.~G.~Russo and A.~A.~Tseytlin,
  ``Large spin expansion of semiclassical 3-point correlators in $\axs$,''
  JHEP {\bf 1102}, 029 (2011)
  [arXiv:1012.2760].
$\bullet$
  C.~Park and B.~Lee,
  ``Correlation functions of magnon and spike,''
  Phys. Rev. D {\bf 83}, 126004 (2011)
  [arXiv:1012.3293].
$\bullet$
  E.~I.~Buchbinder and A.~A.~Tseytlin,
  ``Semiclassical four-point functions in $\axs$,''
  JHEP {\bf 1102}, 072 (2011)
  [arXiv:1012.3740].
$\bullet$
  D.~Bak, B.~Chen and J.~Wu,
  ``Holographic Correlation Functions for Open Strings and Branes,''
  JHEP {\bf 1106}, 014 (2011)
  [arXiv:1103.2024].
$\bullet$
  D.~Arnaudov, R.~C.~Rashkov and T.~Vetsov,
  ``Three- and four-point correlators of operators dual to folded string solutions in $\axs$,''
  Int. J. Mod. Phys. A {\bf 26}, 3403 (2011)
  [arXiv:1103.6145].
$\bullet$
  A.~Bissi, C.~Kristjansen, D.~Young and K.~Zoubos,
  ``Holographic three-point functions of giant gravitons,''
  JHEP {\bf 1106}, 085 (2011)
  [arXiv:1103.4079].
$\bullet$
  R.~Hern\'andez,
  ``Three-point correlators for giant magnons,''
  JHEP {\bf 1105}, 123 (2011)
  [arXiv:1104.1160].
$\bullet$
  X.~Bai, B.~Lee and C.~Park,
  ``Correlation function of dyonic strings,''
  Phys. Rev. D {\bf 84}, 026009 (2011)
  [arXiv:1104.1896].
$\bullet$
  C.~Ahn and P.~Bozhilov,
  ``Three-point Correlation functions of Giant magnons with finite size,''
  Phys. Lett. B {\bf 702}, 286 (2011)
  [arXiv:1105.3084].
$\bullet$
  B.~Lee and C.~Park,
  ``Finite size effect on the magnon's correlation functions,''
  Phys. Rev D {\bf 84}, 086005 (2011)
  [arXiv:1105.3279].
$\bullet$
  D.~Arnaudov and R.~C.~Rashkov,
  ``Quadratic corrections to three-point functions,''
  Fortschr. Phys. {\bf 60}, 217 (2012)
  [arXiv:1106.0859].
$\bullet$
  G.~Georgiou,
  ``SL(2) sector: weak/strong coupling agreement of three-point correlators,''
  JHEP {\bf 1109}, 132 (2011)
  [arXiv:1107.1850].
$\bullet$
  P.~Bozhilov,
  ``More three-point correlators of giant magnons with finite size,''
  JHEP {\bf 1108}, 121 (2011)
  [arXiv:1107.2645].
$\bullet$
  M.~Michalcik, R.~C.~Rashkov and M.~Schimpf,
  ``On semiclassical calculation of three-point functions in $AdS_5\times T^{1,1}$,''
  Mod. Phys. Lett. A {\bf 27}, 1250091 (2012)
  [arXiv:1107.5795].
$\bullet$
  P.~Bozhilov,
  ``Three-point correlators: finite-size giant magnons and singlet scalar operators on higher string levels,''
  Nucl. Phys. B {\bf 855}, 268 (2012)
  [arXiv:1108.3812].
$\bullet$
  A.~Bissi, T.~Harmark and M.~Orselli,
  ``Holographic 3-point function at one loop,''
  JHEP {\bf 1202}, 133 (2012)
  [arXiv:1112.5075].
$\bullet$
  P.~Caputa, R.~Koch and K.~Zoubos,
  ``Extremal vs. Non-Extremal Correlators with Giant Gravitons,''
  JHEP {\bf 1208}, 143 (2012)
  [arXiv:1204.4172].
$\bullet$
  B.~Lee, B.~Gwak and C.~Park,
  ``Correlation functions of the ABJM model,''
  [arXiv:1211.5838].
$\bullet$
  P.~Bozhilov,
  ``Leading finite-size effects on some three-point correlators in $AdS_5\times S^5$,''
  [arXiv:1212.3485].


\bibitem{Klose:2011}
  T.~Klose and T.~McLoughlin,
  ``A light-cone approach to three-point functions in $AdS_5\times S^5$,''
  JHEP {\bf 1204}, 080 (2012)
  [arXiv:1106.0495].
$\bullet$
  S.~Ryang,
  ``Extremal Correlator of Three Vertex Operators for Circular Winding Strings in $AdS_5\times S^5$,''
  JHEP {\bf 1111}, 026 (2011)
  [arXiv:1109.3242].
$\bullet$
  R.~A.~Janik and A.~Wereszczynski,
  ``Correlation functions of three heavy operators: The AdS contribution,''
  JHEP {\bf 1112}, 095 (2011)
  [arXiv:1109.6262].
$\bullet$
  Y.~Kazama and S.~Komatsu,
  ``On holographic three point functions for GKP strings from integrability,''
  JHEP {\bf 1201}, 110 (2012)
  [arXiv:1110.3949].
$\bullet$
  E.~I.~Buchbinder and A.~A.~Tseytlin,
  ``Semiclassical correlators of three states with large $S^5$ charges in string theory in $\axs$,''
  Phys. Rev. D {\bf 85}, 026001 (2012)
  [arXiv:1110.5621].
$\bullet$
  S.~Ryang,
  ``Three-Point Correlator of Heavy Vertex Operators for Circular Winding Strings in $AdS_5\times S^5$,''
  Phys. Lett. B {\bf 713}, 122 (2012)
  [arXiv:1204.3688].
$\bullet$
  Y.~Kazama and S.~Komatsu,
  ``Wave functions and correlation functions for GKP strings from integrability,''
  JHEP {\bf 1209}, 022 (2012)
  [arXiv:1205.6060].
$\bullet$
  J.~Minahan,
  ``Holographic three-point functions for short operators,''
  JHEP {\bf 1207}, 187 (2012)
  [arXiv:1206.3129].


\bibitem{Georgiou:2013}
  G.~Georgiou, B.~Lee and C.~Park,
  ``Correlators of massive string states with conserved currents,''
  [arXiv:1301.5092].


\bibitem{Freedman:1998tz}
  D.~Z.~Freedman, S.~D.~Mathur, A.~Matusis and L.~Rastelli,
  ``Correlation functions in the CFT$_d$/AdS$_{d+1}$ correspondence,''
  Nucl. Phys. B {\bf 546}, 96 (1999)
  [hep-th/9804058].

\bibitem{Chalmers:1998xr}
  G.~Chalmers, H.~Nastase, K.~Schalm and R.~Siebelink,
  ``$R$-current correlators in $N=4$ super-Yang-Mills theory from anti-de Sitter supergravity,''
  Nucl. Phys. B {\bf 540}, 247 (1999)
  [hep-th/9805105].

\bibitem{Fradkin:1998}
  E.~S.~Fradkin and Y.~M.~Palchik,
  ``New developments in D-dimensional conformal quantum field theory,''
  Phys. Rep. {\bf 300}, 1 (1998).

\bibitem{Kim:1985}
  H.~J.~Kim, L.~J.~Romans and P.~van~Nieuwenhuizen,
  ``The Mass Spectrum Of Chiral N=2 D=10 Supergravity On S5,''
  Phys. Rev. D {\bf 32}, 389 (1985).

\bibitem{Osborn:1993cr}
  H.~Osborn and A.~Petkou,
  ``Implications of conformal invariance in field theories for general dimensions,''
  Ann. Phys. {\bf 231}, 311 (1994)
  [hep-th/9307010].

\bibitem{Erdmenger:1996yc}
  J.~Erdmenger and H.~Osborn,
  ``Conserved currents and the energy-momentum tensor in conformally invariant theories for general dimensions,''
  Nucl. Phys. B {\bf 483}, 431 (1997)
  [hep-th/9605009].

\bibitem{Howe:1998zi}
  P.~S.~Howe, E.~Sokatchev and P.~C.~West,
  ``3-point functions in ${\cal N}=4$ Yang-Mills,''
  Phys. Lett. B {\bf 444}, 341 (1998)
  [hep-th/9808162].

\bibitem{Arutyunov:1999nw}
  G.~Arutyunov and S.~Frolov,
  ``Three-point Green function of the stress-energy tensor in the AdS-CFT correspondence,''
  Phys. Rev. D {\bf 60}, 026004 (1999)
  [hep-th/9901121].

\bibitem{Liu:1998bu}
  H.~Liu and A.~A.~Tseytlin,
  ``$D=4$ super Yang-Mills, $D=5$ gauged supergravity and $D=4$ conformal supergravity,''
  Nucl. Phys. B {\bf 533}, 88 (1998)
  [hep-th/9804083].


\bibitem{Escobedo:2010}
  J.~Escobedo, N.~Gromov, A.~Sever and P.~Vieira,
  ``Tailoring Three-Point Functions and Integrability,''
  JHEP {\bf 1109}, 028 (2011)
  [arXiv:1012.2475].
$\bullet$
  J.~Escobedo, N.~Gromov, A.~Sever and P.~Vieira,
  ``Tailoring Three-Point Functions and Integrability II. Weak/strong coupling match,''
  JHEP {\bf 1109}, 029 (2011)
  [arXiv:1104.5501].
$\bullet$
  J.~Caetano and J.~Escobedo,
  ``On four-point functions and integrability in ${\cal N}=4$ SYM: from weak to strong coupling,''
  JHEP {\bf 1109}, 080 (2011)
  [arXiv:1107.5580].
$\bullet$
  O.~Foda,
  ``${\cal N}=4$ SYM structure constants as determinants,''
  JHEP {\bf 1203}, 096 (2012)
  [arXiv:1111.4663].
$\bullet$
  G.~Georgiou, V.~Gili, A.~Grossardt and J.~Plefka,
  ``Three-point functions in planar ${\cal N}=4$ super Yang-Mills Theory for scalar operators up to length five at the one-loop order,''
  JHEP {\bf 1204}, 038 (2012)
  [arXiv:1201.0992].
$\bullet$
  N.~Gromov and P.~Vieira,
  ``Quantum Integrability for Three-Point Functions,''
  [arXiv:1202.4103].
$\bullet$
  G.~Grignani and A.~Zayakin,
  ``Matching Three-point Functions of BMN Operators at Weak and Strong coupling,''
  JHEP {\bf 1206}, 142 (2012)
  [arXiv:1204.3096].
$\bullet$
  G.~Grignani and A.~Zayakin,
  ``One-loop three-point functions of BMN operators at weak and strong coupling,''
  JHEP {\bf 1209}, 087 (2012)
  [arXiv:1205.5279].
$\bullet$
  A.~Bissi, G.~Grignani and A.~Zayakin,
  ``The SO(6) Scalar Product and Three-Point Functions from Integrability,''
  [arXiv:1208.0100].
$\bullet$
  J.~Caetano and J.~Toledo,
  ``$\chi$--Systems for Correlation Functions,''
  [arXiv:1208.4548].

\bibitem{Hirano:2012}
  S.~Hirano, C.~Kristjansen and D.~Young,
  ``Giant Gravitons on $AdS_4\times\mathbb{C}\mathrm{P}^3$ and their Holographic Three-point Functions,''
  JHEP {\bf 1207}, 006 (2012)
  [arXiv:1205.1959].
$\bullet$
  P.~Caputa and B.~Mohammed,
  ``From Schurs to Giants in ABJ(M),''
  JHEP {\bf 1301}, 055 (2013)
  [arXiv:1210.7705].
$\bullet$
  A.~Bissi, C.~Kristjansen, A.~Martirosyan and M.~Orselli,
  ``On Three-point Functions in the $AdS_4/CFT_3$ Correspondence,''
  JHEP {\bf 1301}, 137 (2013)
  [arXiv:1211.1359].

\end{thebibliography}
\end{document}